\documentclass[floatfix,prb,aps,twocolumn,showpacs,preprintnumbers,
amsmath,amssymb]{revtex4}
\usepackage{graphicx}
\usepackage{dcolumn}
\usepackage{exscale}

\usepackage{bbm}  
\usepackage{amsfonts}
\def\shiftleft#1{#1\llap{#1\hskip 0.04em}}
\def\shiftdown#1{#1\llap{\lower.04ex\hbox{#1}}}
\def\thick#1{\shiftdown{\shiftleft{#1}}}
\def\b#1{\thick{\hbox{$#1$}}}
\begin{document}
\title{Theory of light diffusion in disordered media 
       with linear absorption or gain}
\author{A. Lubatsch}
\affiliation{Physikalisches Institut, 
             Universit\"at Bonn, 53115 Bonn, Germany}
\author{J. Kroha}
\affiliation{Physikalisches Institut, 
             Universit\"at Bonn, 53115 Bonn, Germany}
\author{K. Busch}
\affiliation{Institut f\"ur Theorie der Kondensierten Materie,
             Universit\"at Karlsruhe, 76128 Karlsruhe, Germany, and
             Department of Physics and 
             College of Optics \& Photonics: CREOL \& FPCE, 
             University of Central Florida, Orlando, FL 32816}
\begin{abstract}
\noindent
We present a detailed, microscopic transport theory for light 
in strongly scattering disordered systems whose constituent 
materials exhibit linear absorption or gain. Starting from 
Maxwell's equations, we derive general expressions for transport 
quantities such as energy transport velocity, transport mean 
free path, diffusion coefficient, and absorption/gain length.
The approach is based on a fully vectorial treatment of the
generalized kinetic equation and utilizes an exact
Ward identity (WI). While for loss- and gainless media the WI
reflects local energy conservation, the effects of absorption or
coherent gain are implemented exactly by novel, additional terms
in the WI. As a result of resonant (Mie) scattering from the individual
scatterers, all transport quantities 
acquire strong, frequency-dependent renormalizations, which are,
in addition, characteristically modified by absorption or gain.
We illustrate the influence of various experimentally accessible 
parameters on these quanitities for dilute systems. 
The transport theory presented here may set the stage for a theory
of Random Lasing in three-dimensional disordered media.
\end{abstract}

\maketitle

\section{Introduction}
Despite its long and venerable history, light propagation in 
disordered media continues to be a fascinating and intensely
studied topic. In particular, the discovery of the coherent
back scattering peak \cite{Kug84,Alb85,Wol85} has triggered an
intensive search for Anderson localization \cite{And58} of 
light. Clearly, the unambiguous demonstration of Anderson 
localization of electromagnetic radiation in an appropriate 
disordered dielectric medium in which dephasing and interaction 
effects can be neglected, would mark a major triumph of our 
current understanding of wave propagation. \par
In fact, early works reported anomalously low values of the 
diffusion constant in strongly scattering suspensions 
\cite{Alb91,Tig92}. 
However, it was soon realized \cite{Alb91,Tig92,Tig93} that 
these low values of the diffusion constant are associated with the 
resonant (Mie) scattering of individual scatterers. This leads 
to a frequency-dependent dwell-time that has to be added to the 
time-of-flight between succesive scattering events. As a result, 
the energy transport velocity acquires a corresponding renormalization, 
while the transport mean free path remains essentially unchanged
\cite{Tig93,Kro93}. 
Similarly, an analysis of the coherent back scattering peak together
with the dependence of the transmission through disordered semiconductor 
powders on the sample size, suggested a scaling behavior \cite{Abr79} 
consistent with the onset of Anderson localization \cite{Wie97}.
However, a reexamination of these data ignited a heated debate 
as to how to discriminate Anderson localization from absorption 
effects \cite{Sch99,Wie99}. 
Subsequent experiments on similarly strongly scattering semiconductor
powders \cite{Gom99} did not produce evidence of Anderson localization
and the experimental data could be well explained using a recently
developed effective medium theory that incorporates the resonant
scattering effects alluded to above \cite{Bus95,Bus96}. This 
unsatisfactory state of affairs has generated renewed interest in 
determining novel and unambiguous pathways to Anderson localization
of light. \par
One class of highly interesting systems for multiple scattering of
light are disordered dielectric media whose constituent materials
exhibit one or more forms of optical anisotropies. Moreover, most
optical anisotropies exhibit a certain degree of tunability through
external control parameters, thus creating the possibility of a 
tunable disorder. For instance, Faraday-activty in multiple scattering 
systems breaks the time reversal symmetry of Maxwell's equations, 
leading to profound modifications of the coherent back scattering peak 
\cite{Erb93,Len00,Lab02} and in transport the optical analogue of the 
Hall-effect has been observed \cite{Rik96}. Similarly, disordered 
nematic liquid crystals exhibit anisotropic coherent back scattering 
\cite{Sap04} and anisotropic light diffusion \cite{Tig96,Sta96}. 
However, the strength of multiple scattering in bulk nematic liquid 
crystals is determined by the fluctuations of the nematic director 
field and the difference between the liquid crystal's ordinary 
and extraordinary index of refraction. For an observation of Anderson 
localization of light in such systems, it will become necessary to 
enhance the multiple scattering effects, for instance, by infiltrating 
the nematic liquid crystal into the void regions of strongly scattering 
photonic crystals \cite{Bus99}.\par
Equally intriguing is to combine the effects of multiple scattering
of light with optical gain and to investigate how the two phenomena 
mutually influence each other. From the optical gain point of view,
diffusing light will spend more time interacting with active material
in a characteristic volume than ballistically propagating light. 
If this interaction time exceeds
the (spontaneous) decay time of the active material, avalanche-like
intensity bursts, induced by {\it incoherent} feedback, could occur. 
In fact, early theoretical 
work \cite{Let68} suggested this very possibility and has recently 
been observed \cite{Law94}. 
From a wave propagation point of view, optical gain increases the 
relative weight of long trajectories in the sample and, therefore, 
will modify a number of wave interference effects such as coherent 
back scattering \cite{Wie95} and possibly Anderson localization. 
Very recent experiments point
to the possibility that such long trajectories provide a feedback
mechanism which leads to modes with narrow laser-like emission lines 
that extend across the entire sample \cite{Muj04}. However, the 
coherence properties of the emitted light still need to be analyzed
in order to establish that laser action is indeed taking place in these
systems. True laser action from localized regions in disordered 
dielectric samples with optical gain, termed Random Lasing, 
has been observed \cite{Cao99,Cao00,Fro99}
earlier, where measurements of the photon statistics of the emitted 
light \cite{Cao01} have unambiguously demonstrated a {\it coherent} 
laser feedback mechanism.
It has been discussed in the form of varying degrees of coupling between 
so-called quasi-states \cite{Jia00,Cao03a}.
For a recent review of multiple 
scattering in amplifying media, we refer the reader to Ref. 
\cite{Cao03b}. 
Random Lasing has potential applications ranging from micro-lasers and
optical fingerprint markers \cite{Wie00} 
to the detection of cancerous tissue \cite{Pol04}. In addition, we
want to note that recently, electrically pumped Random Laser
action has been achieved in Nd-doped powders \cite{Li02,Red04}, 
thus creating potential applications of these systems in omnidirectional 
lighting devices and displays.

Despite these exciting developments, a convincing connection between
Random Lasing and Anderson localization of light has not been 
demonstrated as of yet. This may be attributed to the fact, that to 
date the theory of random amplifying media either employs purely 
numerical methods in one spatial dimension \cite{Jia00} or treats the 
multiple scattering part within certain approximation schemes that 
cannot account for the interference effects that lead to Anderson 
localization.
These schemes include modeling the electromagnetic wave propagation 
through diffusion equations for the intensity \cite{Wie96,Flo04b} or
within the so-called ladder-approximation of the Bethe-Salpeter
equation \cite{Flo04a}. 
As a result, it is unclear, whether 
Anderson localization of light
is a necessary condition for true Random Lasing (coherent feedback) 
nor whether the modified coherence properties in the 
lasing state have, in turn, an influence on the transition from the
diffusion regime to the Anderson localized regime itself.
In fact, the very concept of describing the Anderson localization 
transition in terms of a vanishing diffusion coefficient as an 
order parameter, familiar from systems with energy or particle number
conservation, becomes questionable in dissipative or active media,
where another channel for change of the energy density exists 
besides diffusion. 

In the present paper, we report our progress towards 
answering these questions.  We develop a fully vectorial transport theory 
for multiple scattering of light in random media whose constituent 
materials are isotropic and exhibit linear absorption or gain.
In Section II, we derive the tensorial kinetic
equation for the intensity correlation function of electromagnetic radiation. 
The conservation of energy in media without loss or gain is incorporated
in a field theoretical way by means of an exact Ward identity (WI), the
effects of loss, gain as well as frequency dependence of the material 
parameters being represented by additional terms. The proof of this generalized
WI is presented in Section III.
Subsequently, we solve the kinetic equation in the hydrodynamic limit
by formulating, with the help of the WI, the continuity equation in Section
IV and Fick's law in Section V, respectively. These equations relate the 
energy density correlation tensor and the energy current correlation tensor 
to each other. They allow to identify, in the hydrodynamic limit, exact expressions 
for quantities like the energy transport velocity, the transport mean free path
and the absorption/gain length in terms of the 
irreducible single-photon self-energy and the two-photon irreducible vertex.
We recover the well-known Mie scattering renormalization term to the transport
velocity \cite{Tig93,Kro93}, albeit in vectorial form, along with additional, 
characteristic renormalizations originating from absorption or gain.
Section VI features numerical results for the energy velocity in dilute
systems of spherical scatterers, where all quantities can be evaluated within the 
independent scattering approximation. The Mie resonance dips in the energy transport
velocity acquire a characteristic, absorption-induced broadening and
gain-induced narrowing, which may be interpreted as due to the reduction (enhancement)
of higher-order scattering contributions by absorption (gain).
In contrast to previous approaches, our theory can be
extended to systematically include wave interference effects 
(``Cooperon'' contributions) and, thus, to 
address the Anderson localization transition by 
virtue of a self-consistent extension \cite{Vol80a,Vol80b,Kro90,Kro93} 
together with a replacement of the linear absorption/gain through a coupling
to the master equation of the active medium \cite{Flo04a}.

\section{Basic theory of electromagnetic wave transport}
We consider the propagation of light in a system of randomly 
positioned scatterers with isotropic dielectric constant 
$\epsilon_s$ immersed in a host medium with isotropic dielectric 
constant $\epsilon_h$. In addition, we allow for the possibility
of having absorption or amplification in both the scatterer material
and the host medium by ascribing an imaginary part to the respective
dielectric constants. By virtue of the Kramers-Kronig relations 
between real and imaginary parts of the dielectric constant, we are 
required to consider materials with complex frequency dependent 
dielectric constant $\epsilon_s \equiv \epsilon_s(\omega)$ and 
$\epsilon_h \equiv \epsilon_h(\omega)$. The resulting dielectric
constant 
\begin{equation}
\label{eq:wave}
\epsilon(\vec{r},\omega) = \epsilon_h + 
                           \left( \epsilon_s - \epsilon_h \right)
                           V(\vec{r}),
\end{equation} 
describes the arrangement of scatterers through the function
$V(\vec{r}) = \sum_{\vec{R}} S_{\vec{R}}(\vec{r}-\vec{R})$ which 
consists of a set of localized shape functions $S_{\vec{R}}(\vec{r})$
of the invididual scatterers at random locations $\vec{R}$.
The corresponding wave equation for a time harmonic wave with 
frequency $\omega$ and amplitude $E_\omega(\vec{r})$ reads
\begin{equation}
\label{eq:field}
\vec{\nabla} \times \vec{\nabla} \times \vec{E}_\omega( \vec{r})
- \frac{\omega^2}{c^2} \, \epsilon (\vec{r},\omega) 
  \vec{E}_\omega(\vec{r})
= \omega \, \vec{J}_\omega(\vec{r}).
\end{equation}
Here, $c$ denotes the vacuum speed of light, and 
$\vec{J}_\omega (\vec{r}) = (4\pi i/c^2)\, \vec{j}_\omega (\vec{r})$ represents an 
excitation of the wave field through an external current source $\vec{j}$.
The dielectric function $\epsilon (\vec{r},\omega)$ carries a 
positive or negative imaginary part for absorbing or coherently
amplifying media, respectively. 
\par
For a specific realization $\epsilon(\vec{r},\omega)$ of disorder,
the formal solution to Eq. (\ref{eq:wave}) is given in terms of the
Green's tensor (of 2nd rank in the spatial vector components) 
$\b{G}(\vec{r}, \vec{r}^\prime,\omega)$ as
\begin{equation}
\label{eq:single-gf}
\vec{E}_\omega(\vec{r}) 
        = \int d^3 r^\prime  \, 
          \b{G}(\vec{r}, \vec{r}^\prime,\omega) \, 
          {\vec{J}}_\omega(\vec{r}^\prime).
\end{equation}
For the analytical developments as well as for the comparison with experiments 
it is advantageous to calculate disorder-averaged quantities. Since these are
translationally invariant, the transport theory must be formulated for
correlation functions rather than single-particle properties. Denoting the 
average over disorder realizations by $\langle \dots \rangle$, and 
changing to a Fourier representation of averaged quantities,
the disorder averaged 
Green's tensor $\b{G}^\omega(\vec{r} - \vec{r}^\prime) \equiv 
 \langle \b{G}(\vec{r}, \vec{r}^\prime,\omega) \rangle$ and its Fourier
transform $\b{G}^{\omega}_{\vec{k}}$ may be expressed in terms
of the free Green's tensor $\b{G}_{0}(\vec{k},\omega)$ in the background
medium,
\begin{equation}  
\b{G}_{0}(\vec{k},\omega) = 
\left( \epsilon_h(\omega) (\omega^2/c^2) - 
       \vert \vec{k} \vert^2 \right)^{-1}
\left( \mathbbm{1}-\hat{k}\hat{k}^T \right),
\end{equation}
and the self-energy tensor $\b{\Sigma}^{\omega}_{\vec{k}}$, which 
represents the effects of scattering by the random perturbation 
or ``scattering potential'',  
$(\omega^2/c^2)(\epsilon_s(\omega) - \epsilon_h(\omega))V(\vec{r})$, as
\begin{equation}
\label{gf}
\b{G}^{\omega}_{\vec{k}} = \left( \b{G}^{-1}_0(\vec{k},\omega) -
                                         \b{\Sigma}^{\omega}_{\vec{k}}
                                  \right)^{-1}.
\end{equation}
In the above expressions, we have introduced the three-dimensional 
unit tensor $\mathbbm{1}$ and the dyadic product $\hat{k}\hat{k}^T$ 
of unit vectors $\hat{k}$ in the direction of $\vec{k}$. 
Throughout this paper the propagators $\b{G}$ are understood as the 
retarded ones and complex conjugated propagators $\b{G}^*$ as the 
advanced ones.  
For practical calculations, the self-energy tensor
$\b{\Sigma}^{\omega}_{\vec{k}}$ has to be evaluated within consistent
approximations such as the independent scatterer approximation (see
section VI) or the Coherent Potential Approximation 
\cite{Kro90,Gon92}.\par
\begin{figure} 
\begin{center}
\rotatebox{0}{\scalebox{0.7}{\includegraphics{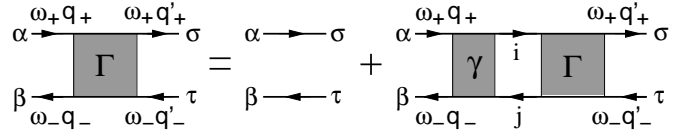}}}
\end{center}
\caption{ 
Diagrammatic representation  of the Bethe - Salpeter Equation, 
Eq. (\ref{Gamma}), indicating the
tensorial structure as well as  the full index notation.                     
}
\label{plot1_fig}
\end{figure}
The disorder-averaged field correlation tensor is definied as 
\begin{eqnarray}
{\b{I}}(\vec{r}_1,\vec{r}_2; \omega_1, \omega_2) & = &
\langle \vec{E}_{\omega_1}(\vec{r}_1) \, \vec{E}^{*,T}_{\omega_2}(\vec{r}_2)]
\rangle 
\nonumber \\
& = & \! \! \int d^3 r_3 \! \! \int  d^3 r_4 \, \,
           \b{\Gamma}_{\omega_1,\omega_2}(\vec{r}_1, \vec{r}_2, \vec{r}_3,\vec{r}_4) \nonumber \\
&   &      \qquad \times \b{S}_{\omega_1,\omega_2}(\vec{r}_3, \vec{r}_4),
\end{eqnarray}
where the system's (averaged) correlation tensor (of fourth rank) 
$\b{\Gamma}_{\omega_1,\omega_2}(\vec{r}_1,\vec{r}_2,\vec{r}_3,\vec{r}_4)$
is independent of the source correlation tensor 
$\b{S}_{\omega_1,\omega_2}(\vec{r}_3, \vec{r}_4) = 
\vec{J}_{\omega_1}(\vec{r}_3) \vec{J}^{*,T}_{\omega_1}(\vec{r}_4)$ and 
is given in terms of the disorder averaged tensor product of 
(unaveraged) Green's tensors according to
\begin{eqnarray} 
\b{\Gamma}_{\omega_1,\omega_2}(\vec{r}_1,\vec{r}_2,\vec{r}_3,\vec{r}_4) 
& = & \nonumber \\
& & \! \! \! \! \! \! \! \! \! \! \! \! \! \! \! \! \!
\langle        \b{G}(\vec{r}_1,\vec{r}_3,\omega_1)   
       \otimes \b{G}^*(\vec{r}_2,\vec{r}_4,\omega_2)
\rangle \ ,
\end{eqnarray}
where $( . \otimes . )$ denotes the tensor product of two 2nd rank tensors
operating in the space of retarded and advanced propagators, respectively. 
Similar to the disorder averaged Green's tensor 
$\b{G}^{\omega}_{\vec{k}}$, we introduce the spatial Fourier
transform $\b{\Gamma}^\omega_{\vec{q} \vec{q}^\prime}(\vec{Q},\Omega)$
of the correlation tensor
\begin{equation}
\label{corr}
\b{\Gamma}^\omega_{\vec{q} \vec{q}^\prime}(\vec{Q},\Omega)
 = \langle \b G_{\omega_+}(\vec{q}_+,\vec{q}^\prime_+)   
       \otimes \b G^*_{\omega_-}(\vec{q}_-,\vec{q}^\prime_-) \rangle,
\end{equation} 
where the transition to center of mass and relative frequencies,
$\Omega$, $\omega$, and momenta, $\vec Q$, $\vec q$, respectively, 
with $\omega_{1,2} = \omega \pm \Omega /2\equiv \omega_{\pm}$, 
$\vec{q}_{1, 2} = \vec{q} \pm \vec{Q}/2\equiv \vec q_{\pm}$ and 
$\vec{q}_{3, 4} = \vec{q}^\prime \pm \vec{Q}/2\equiv \vec q_{\pm}$, 
facilitates an investigation of the correlation tensor's long-time 
($\Omega \to 0$) and long-distance ($ \vert \vec{q} \vert \to 0$) 
behavior. $\Omega$, $\vec Q$ are associated with the time and 
position dependence of the electromagnetic energy density in the
system, while $\omega$ represents the frequency of light.\par
$\b{\Gamma}^\omega_{\vec{q} \vec{q}^\prime}(\vec{Q},\Omega)$ can be expressed 
in terms of the irreducible vertex tensor 
$\b{\gamma}^{\omega}_{\vec{q} \vec{q}^{\prime}}(\vec{Q},\Omega)$, the
two-photon analogue of the self-energy tensor, via the 
Bethe-Salpeter equation
\begin{eqnarray}
\b{\Gamma}^\omega_{\vec{q} \vec{q}^\prime}(\vec{Q},\Omega) =
\b{G}^{\omega_+}_{\vec{q}_+} \otimes \left(\b{G}^{\omega_-}_{\vec{q}_-} \right)^*
\left[ (2 \pi)^3 \delta(\vec{q}-\vec{q}^\prime) \mathbbm{1}\otimes\mathbbm{1}  \right.
\nonumber \\ 
  + \left.
    \int \frac{d^3 q^{\prime \prime}}{(2 \pi)^3} \,
    \b{\gamma}^\omega_{\vec{q} \vec{q}^{\prime \prime}}
               (\vec{Q},\Omega) 
    \,
    \b{\Gamma}^\omega_{\vec{q}^{\prime \prime} \vec{q}^\prime}
               (\vec{Q},\Omega)
 \right] \ .
\label{Gamma}
\end{eqnarray}
In Eq. (\ref{Gamma}), both the irreducible vertex tensor 
$\b{\gamma}^{\omega}_{\vec{q} \vec{q}^{\prime}}(\vec{Q},\Omega)$ and
the correlation tensor 
$\b{\Gamma}^\omega_{\vec{q} \vec{q}^\prime}(\vec{Q},\Omega)$ are tensors
of fourth rank which operate both in retarded and advanced space (see
Eq. (\ref{corr})). The notation 
$\b{\gamma}^{\omega}_{\vec{q} \vec{q}^{\prime}}(\vec{Q},\Omega) \,
\b{\Gamma}^\omega_{\vec{q} \vec{q}^\prime}(\vec{Q},\Omega)$ 
in Eq. (\ref{Gamma}) 
implies contraction of both, retarded and advanced, intermediate indices,
\begin{eqnarray}
\left[\b\gamma^\omega_{\vec{q} \vec{q}^{\prime \prime}}(\vec{Q},\Omega) 
      \,
      \b\Gamma^\omega_{\vec{q}^{\prime \prime} \vec{q}^\prime}
                      (\vec{Q},\Omega)\right]_{\alpha\beta\sigma\tau}
      & = & \nonumber \\
& & 
\! \! \! \! \! \! \! \! \! \! \! \! \! \! \! \! \! \! \! \! \! \!
\! \! \! \! \! \! \! \! \! \! \! \! \! \! \! \! \! \! \! \! \! \!
\! \! \! \! \! \! \! \! \! \! \! \! \! \! \! \! \! \! \! \!
\left[\b\gamma^\omega_{\vec{q} \vec{q}^{\prime \prime}}
               (\vec{Q},\Omega)\right]_{\alpha \beta i  j}
 \left[\b\Gamma^\omega_{\vec{q}^{\prime \prime} \vec{q}{^\prime}}
               (\vec{Q},\Omega)\right] _{i j \sigma \tau},
\end{eqnarray}
where summation over repeated indices is implied,
compare  Fig.\ref{plot1_fig}.
For tensor products of 2nd rank tensors 
in retarded and advanced space,  
$\b{B},\b{C}$ and $\b{E},\b{F}$, respectively, we have the multiplication
rule
\begin{equation}
({\b B}\otimes {\b E}) \;\;  ({\b C}\otimes {\b F})
=
({\b B}{\b C}) \otimes ({\b E}  {\b F}),
\end{equation}
where ${\b B}{\b C}$ and ${\b E}{\b F}$ denotes the standard
(matrix) product between tensors of second rank.
It follows the general identity
\begin{eqnarray}
\label{2.11}
{\b G}^{{\omega_{+}}}_{{\vec{q}}_{+}} 
\otimes
\left(
{\b G} ^{{\omega_{-}}}_{{\vec{q}}_{-}} 
\right)^* =
\left( 
{\b G}^{{\omega_{+}}}_{{\vec{q}}_{+}} \otimes \mathbbm{1}
- 
\mathbbm{1}\otimes
\left({\b G}^{{\omega_{-}}}_{{\vec{q}}_{-}} \right)^*
\right)
\nonumber \\
\times
\left(
\mathbbm{1}\otimes
\left[
\left({\b G}^{{\omega_{-}}}_{{\vec{q}}_{-}} \right)^*\right
]^{-1}
-
\left[{\b G}^{{\omega_{+}}}_{{\vec{q}}_{+}}\right]^{-1}
\otimes \mathbbm{1}
\right)^{-1},
\end{eqnarray}
which allows us, after integration over the 
momentum $\vec{q}^\prime$, to rewrite the Bethe-Salpeter equation (\ref{Gamma}) 
as a kinetic equation (a generalized Boltzmann equation)
for the system's integrated intensity correlation tensor
\begin{equation}
\b{\Phi}^{\omega}_{\vec{q}} (\vec{Q}, \Omega) = 
 \int \frac{d^3 q^\prime}{(2 \pi)^3}  \, 
      \b{\Gamma}^{\omega}_{\vec{q} \vec{q}^\prime}(\vec{Q},\Omega).
\end{equation}
Explicitly, the generalized Boltzmann equation reads,
\begin{eqnarray}
\label{boltzmann}
\left[
-\Delta  {g}_{\omega}\mathbbm{1}\otimes\mathbbm{1}
-\Delta\hat{\b L}_{\vec{q}}
+\Delta{\b \Sigma}^{\omega}_{\vec{q}}
\right]  \,
{\b\Phi}^{\omega}_{\vec{q}}
& = & \nonumber \\ 
& & \hspace*{-4.5cm}
\Delta {\b G}^{\omega}_{\vec{ q}} 
\left[
 \,\mathbbm{1}\otimes\mathbbm{1}
+
\int\frac{{\rm d}^3 q^{\prime\prime} }{(2\pi)^3} 
{\b \gamma}^{\omega}_{\vec{q}{\vec{ q}^{\prime\prime}}} \,
{\b\Phi}^{\omega}_{\vec{q}^{\prime\prime}}
\right]. 
\end{eqnarray}
Here, we have introduced short-hand notations for certain
differences and, for later use, sums of 4th rank tensors,
\begin{eqnarray}
\label{define_the_Deltas}
\Delta  {g}_{\omega}(\Omega) 
   & = &   
g(\omega_+) - g^*(\omega_-) \\
\Delta\hat{\b L}_{\vec{q}}({\vec{Q}}) 
   & = &   
\hat{L}({\vec{q}}_+) \otimes \mathbbm{1}  - \mathbbm{1} \otimes 
\left( \hat{L}(\vec{q}_-) \right)^*\\
\Delta{\b \Sigma}^{\omega}_{\vec{q}}(\vec{Q},\Omega)
   & = &
{\b \Sigma}^{\omega_+}_{\vec{q}_+} \otimes \mathbbm{1} -
\mathbbm{1} \otimes \left( {\b \Sigma}^{\omega_-}_{\vec{q}_-}\right)^*\\
\Delta {\b G}^{\omega}_{\vec{q}}(\vec{Q},\Omega) 
   & = &
{\b G}^{\omega_+}_{\vec{q}_+} \otimes \mathbbm{1} -
\mathbbm{1} \otimes \left({\b G}^{\omega_+}_{\vec{q}_+}\right)^*\\
\label{define_the_Pis}
{\b\Pi}{\b \Sigma}^{\omega}_{{\vec{q}}}({\vec{Q}};\Omega)
&=&{\b \Sigma}^{\omega_+}_{\vec{q}_+} 
\otimes \mathbbm{1}
+ 
\mathbbm{1}\otimes
\left({\b
\Sigma}^{\omega_-}_{\vec{q}_-}\right)^*\\
{\b \Pi}{\b G}^{\omega}_{{\vec{q}}}({\vec{Q}};\Omega)
&=&
{\b G}^{\omega_+}_{\vec{q}_+} 
\otimes\mathbbm{1} 
+ 
\mathbbm{1}\otimes
\left({\b G }^{\omega_-}_{\vec{q}_-}\right)^* 
\end{eqnarray}
which employ the definitions 
$g(\omega) = (\omega^2/c^2) \epsilon_h(\omega)$ and
$\hat{L}(\vec{r}) = \vec{\nabla}\times\vec{\nabla}\times$ 
with corresponding Fourier representation 
$\hat{L}(\vec{q})=- \vec{q}\times\vec{q}\times$.\par
The physical interpretation of the generalized Boltzmann equation, 
Eq. (\ref{boltzmann}) starts with the long-time limit, i.e. small $\Omega$
expansion of the first term on the left-hand side (l.h.s.),
$\Delta g_\omega(\Omega) = i g^{(0)}_\omega + 
                             g^{(1)}_\omega \Omega + ...$, where
\begin{eqnarray}
g^{(0)}_\omega &=& 2 \frac{\omega^2}{c^2}  \mbox{Im} \left[  \epsilon_h(\omega)  \right]\\
g^{(1)}_\omega &=& \mbox{Re} \left[\partial_{\omega} \frac{\omega^2}{c^2} \epsilon_h(\omega)  
\right].
\end{eqnarray}
It is seen that the term linear in $\Omega$, $g^{(1)}_\omega$ corresponds
to a rate of change of the correlation tensor 
$\b{\Phi}^{\omega}_{\vec{q}} (\vec{Q}, \Omega)$ (term of O($\Omega$)), while the 
term of $O(\omega ^0)$ describes absorption or emission by the host medium. 
The latter is non-vanishing only if ${\rm Im} \epsilon _h (\omega) \neq 0$.
Similarly, the second term on the l.h.s. represents a drift term
(first order in $\vec{q}$) for 
$\b{\Phi}^{\omega}_{\vec{q}} (\vec{Q}, \Omega)$. 
The third term on the l.h.s of Eq. (\ref{boltzmann}),
$\Delta{\b \Sigma}^{\omega}_{\vec{q}}(\vec{Q},\Omega)$, embodies
single-particle scattering from the external random perturbation, 
while the terms on the right-hand side (r.h.s.) 
represent an effective two-particle collision integral induced by 
disorder averaging. 

The physical
interpretation of the generalized Boltzmann equation suggests the
subsequent strategy for obtaining the solution in the hydrodynamic limit
($\Omega\to 0$, $\vec Q\to 0$), where the collective mode dynamics are governed by
the conservation laws of the system. The electromagnetic wave 
equation (\ref{eq:field}) being 2nd order in time implies (in a loss- and
gainless medium) the local conservation of the energy density rather than
the intensity \cite{Kro93}. Consequently, we will seek solutions for the 
energy density correlation tensor $\b P$ and the energy current
density correlation tensor $\b J$, and recast the kinetic equation in terms 
of these quantities. This is possible because, as seen in section IV,
$\b P$ and $\b J$ are the leading coefficients of the correlator 
${\b\Phi}^{\omega}_{\vec{q}} (\vec Q, \Omega)$ in a small $\vec Q$, $\Omega$
expansion. Before we develop this solution in Sections IV and V, we 
will derive in the next section an exact WI for vector waves, which
relates the photon selfenergy ${\b \Sigma}$ and the irreducible 
two-photon vertex ${\b gamma}$ to each other. It embodies local energy
conservation as well as disspation or gain induced deviations 
in a field theoretical way.

\section{Ward identity}
The derivation of the WI for disordered electronic
systems has been demonstrated for the first time by Vollhard
and W\"olfle \cite{Vol80a,Vol80b} using a diagrammatic approach.
A corresponding WI for scalar classical waves has been
derived, using an algebraic approach, by Barabanenkov et al. 
\cite{Bar91} and, using a diagrammatic technique, by Kroha et al. 
\cite{Kro93}. Barabanenkov et al. have generalized their derivation 
to electromagnetic waves \cite{Bar95a,Bar95b}.
Subsequently, the correct form of this WI has been the 
subject of heated debate \cite{Nie98,Bar01,Nie01}, where a 
consensus has been reached in Ref.\ \onlinecite{Nie01}. However, 
to date, frequency dependent and/or complex dielectric 
functions have not been included in the derivation of WI
for classical waves. Neither have all the implications of the WI
on the renormalization of transport quantities been 
discussed.\par
Therefore, the aim of this section is to derive a WI
for electromagnetic waves in the presence of frequency dependent, 
complex dielectric functions. In the following proof, we choose to 
follow the approach by Barabanenkov et al. \cite{Bar95b}.
A detailed discussion as to how
the WI affects the various transport properties is
postponed to sections IV and  V. \par
We start from the Green's tensor before impurity averaging whose equation of 
motion (see Eqs. (\ref{eq:field}) and (\ref{eq:single-gf})) we
write as
\begin{eqnarray}
\label{define_split_G_Ward_1}
\left(
\hat{\b L}(\vec{r}_1) +  g(\omega_1)
- h(\omega_1)V(\vec{r}_1)  \right)
\\ \nonumber
 \times
\b{\tilde{G}}( \vec{r}_1,\vec{r}_2,\omega_1)
=   \delta(\vec{r}_1 -\vec{r}_2 )\mathbbm{1}.
\end{eqnarray}
Here, we have introduced the, in general, complex quantity 
\begin{eqnarray}
h(\omega) = -(\epsilon_s(\omega)-\epsilon_h(\omega))(\omega^2/c^2)\ .
\end{eqnarray}
Multiplying Eq. (\ref{define_split_G_Ward_1}) with 
$h(\omega_2)^* {\b{\tilde{G}}}^*( \vec{r}_3,\vec{r}_4,\omega_2)$ 
within the appropriate tensor subspace yields
\begin{eqnarray}
\label{MOD_1}
&\Big[&
h^*(\omega_2)
\hat{\b L}(\vec{r}_1)\!\otimes\!\mathbbm{1}
+ 
g(\omega_1)h^*(\omega_2) \mathbbm{1}\!\otimes\!\mathbbm{1}
\\ \nonumber &&\qquad\qquad - 
h(\omega_1)h^*(\omega_2)V(\vec{r}_1)\mathbbm{1}\!\otimes\!\mathbbm{1}
\Big] 
\\ \nonumber &&\qquad\qquad \times
\tilde{\b G}(\vec{r}_1,\vec{r}_2;\omega_1)
\!\otimes\!
\tilde{\b G}^*(\vec{r}_3,\vec{r}_4;\omega_2)
\\ \nonumber &&=
h^*(\omega_2)
\delta(\vec{r}_1-\vec{r}_2)
\mathbbm{1}\!\otimes
\tilde{\b G}^*(\vec{r}_3,\vec{r}_4;\omega_2).
\end{eqnarray}
In the time reversed case, i.e. starting with the complex conjugated
equation, an analogous procedure leads to 
\begin{eqnarray}
\label{MOD_2}
&\Big[&
h(\omega_1)
\mathbbm{1}\!\otimes\! \hat{\b L}^*(\vec{r}_3)
+ 
g(\omega_2)h(\omega_1) \mathbbm{1}\!\otimes\!\mathbbm{1}
\\\nonumber  &&\qquad\qquad    - 
h(\omega_1)h^*(\omega_2)V(\vec{r}_3)\mathbbm{1}\!\otimes\!\mathbbm{1}
\Big]
\\ &&\qquad\qquad  \times \nonumber
\tilde{\b G}(\vec{r}_1,\vec{r}_2;\omega_1)
\!\otimes\!
\tilde{\b G}^*(\vec{r}_3,\vec{r}_4;\omega_2)
\\\nonumber && =
h(\omega_1)
\delta(\vec{r}_3-\vec{r}_4)
\tilde{\b G}(\vec{r}_1,\vec{r}_2;\omega_1)\!\otimes\! \mathbbm{1}.
\nonumber
\end{eqnarray}
Upon substracting Eqs. (\ref{MOD_1}) and Eq. (\ref{MOD_2}),
followed by averging over disorder and taking the limit
of $\vec{r}_1 \to \vec{r}_3$, we combine retarded and advanced 
quantities to what Barabenenkov et al. \cite{Bar95b} refer 
to as the pre-Ward identity.
\begin{eqnarray}
\label{Pre_ward}
\nonumber
\lefteqn{
0=
\lim\limits_{ \vec{r}_1 \rightarrow \vec{r}_3}
\Bigg\{
\Big(
\left[
h^*(\omega_-)  \hat{\b L}(\vec{r}_1)\!\otimes\!\mathbbm{1}
-
h(\omega_+)     \mathbbm{1}\!\otimes\! \hat{\b L}^*(\vec{r}_3)
\right]
}
\\&&
+
\left[
g(\omega_+)h^*(\omega_-)-g^*(\omega_-)h(\omega_+)   
\right] \mathbbm{1}\!\otimes\!\mathbbm{1}
\Big)
\\\nonumber &&\quad 
\langle
\tilde{\b G} (\vec{r}_1,\vec{r}_2;\omega_+)
\otimes
\tilde{\b G}^*(\vec{r}_3,\vec{r}_4;\omega_-)
\rangle
\nonumber
\\&&
-
\langle
h^*(\omega_-)\delta(\vec{r}_1-\vec{r}_2)
\mathbbm{1}\!\otimes
\tilde{\b G}^*(\vec{r}_3,\vec{r}_4;\omega_-)
\nonumber \\&&-
h(\omega_+)\delta(\vec{r}_3-\vec{r}_4)
\tilde{\b G}(\vec{r}_1,\vec{r}_2;\omega_+)\otimes\!\mathbbm{1}
\rangle
\Bigg\}
\nonumber
\end{eqnarray}
The final step in obtaining the WI is to transfer the
pre-Ward identity to momentum space and to cancel or simplify several 
terms with the help of both the Bethe-Salpeter equation and the 
generalized Boltzmann equation (for details we refer to the work
of Barabanenkov et al. \cite{Bar95b}). \par
This finally yields the Ward identity for electromagnetic waves
with complex frequency dependent dielectric functions,
\begin{eqnarray}
\label{Ward_2}
\nonumber
\lefteqn{
\Delta{\b \Sigma}^{\omega}_{{\vec{q}}}({\vec{Q}};\Omega)
-
\int \frac {{\rm d}^3{q}^{\prime}}{(2\pi)^3} \,
\Delta {\b G}^{\omega}_{{\vec{q}}^{\prime}}({\vec{Q}};\Omega) \, 
{\b \gamma}^{\omega}_{{\vec{q}}^{\prime}{{\vec{q}}}}({\vec{Q}},\Omega)
} \\
&& =
\frac {h(\omega_+)-h^*(\omega_-)}
{h(\omega_+)+h^*(\omega_-)}
\Bigg[
{\b \Pi}{\b \Sigma}^{\omega}_{{\vec{q}}}({\vec{Q}};\Omega)
    \\&& +
\int \frac {{\rm d}^3{q}^{\prime}}{(2\pi)^3} \,
{\b \Pi} {\b G}^{\omega}_{{\vec{q}}^{\prime}}({\vec{Q}};\Omega) \, 
{\b \gamma}^{\omega}_{{\vec{q}}^{\prime}{{\vec{q}}}}({\vec{Q}},\Omega)
\Bigg].
\nonumber
\end{eqnarray}
As compared to the case of electronic wave propagation in a disordered
solid \cite{Vol80a,Vol80b}, the non-zero r.h.s. of Eq. (\ref{Ward_2}) 
constitutes a novel term which originates from the dependence 
of the ``scattering potential'' 
$(\omega^2/c^2) (\epsilon _s(\omega) - \epsilon _h (\omega)) V(\vec{r})$  
on the light frequency as well as from its possible imaginary part, 
and will lead below to
a renormalization of the energy transport velocity \cite{Alb91,Tig93,Kro93}.
In fact, in the case of frequency independent and real dielectric
constant, we have that the prefactor of this term,
$(h(\omega_+)-h^*(\omega_-))/(h(\omega_+)+h^*(\omega_-)) = 
 4 \omega \Omega/( 4 \omega^2 + \Omega^2)$, takes on a form that has 
been discovered earlier \cite{Kro93}. Since in this case the r.h.s. 
of Eq.\ (\ref{Ward_2})is $\propto \Omega$ it leads to a renormalization
of the energy transport velocity \cite{Kro93}.
However, in the present
case of absorptive or amplifying media, the r.h.s.  
contributes also in zero-th order in $O\Omega$, signaling that
absorption and gain induce more severe effects than 
renormalizing the energy transport velocity, namely a mass term 
for the diffusion modes, see below.

\section{Continuity equation}
We now proceed with solving the kinetic equation (\ref{boltzmann}). 
From the 2nd order wave equation (\ref{eq:field}) it follows that 
for a homogeneous system ($\epsilon (\vec{r}) = \epsilon _h =const.$)
the quantities
\begin{eqnarray} 
\rho _E &=& i \frac{\sqrt{\epsilon _h}}{c} 
\left( 
\vec{E} ^{\dagger}\cdot \frac{\partial \vec{E}}{\partial t} - 
\frac{\partial \vec{E}^{\dagger}}{\partial t}\cdot \vec{E}
\right) \\
\vec{j}_E &=& i  \frac{c}{\sqrt{\epsilon _h}} 
\left( 
\vec{E} ^{\dagger}\cdot (\nabla \vec{E}) - 
(\nabla \vec{E}^{\dagger})\cdot \vec{E} 
\right) 
\end{eqnarray}
obey the continuity equation $\partial \rho _E / \partial t +
\nabla \cdot \vec{j}_E =0$, and may be interpreted as 
energy density and energy current density, respectively, 
where $\bar{c} = c/\sqrt{{\rm Re}\epsilon _h (\omega )}$ 
is the phase velocity of the homogeneous medium. 
In order to obtain a similar relation for correlation functions
in a random medium we, therefore, define the energy density--energy density 
correlation tensor $ {\b P}^{\omega}_{\mbox{\tiny E}}$,
and the longitudinal energy current--energy density  correlation tensor
$ {\b J}^{\omega}_{\mbox{\tiny E}}$ in Fourier representation as
\begin{eqnarray}
\label{PE}
{\b P}^{\omega}_{\mbox{\tiny E}}(\vec{Q}, \Omega)
 &\!\! =\!\! & 
      \omega^2  \b c^{-2}_{\mbox{\tiny P}} 
     \!\! \int\!\!\! \frac{d^3q}{(2 \pi)^3} 
      \,\b \Phi^{\omega}_{\vec{q}} (\vec{Q}, \Omega), \\ 
\label{JE}
{\b J}^{\omega}_{\mbox{\tiny E}}(\vec{Q}, \Omega) &\! \!= \!\! & 
\omega {\b v}_{\mbox{\tiny E}} (\omega) 
              {\b c}^{-1}_{\mbox{\tiny P}} 
\!\!\int\!\!\! \frac{d^3q}{(2 \pi)^3} \, 
      (\vec{q} \cdot \hat{Q}) \, \b \Phi^{\omega}_{\vec{q}} 
(\vec{Q}, \Omega),
\end{eqnarray}
where
\begin{eqnarray}
\label{phasevelocity}
{\b c}_{\mbox{\tiny P}} &\!\! = \!\!& \bar{c}^2
        \left[ \mathbbm{1}-\mbox{Re}(\b\Sigma^{\omega}_{\vec{k} = \omega/c})/
                     (\omega^2/c^2)\right]^{-1} 
\end{eqnarray}
defines the averaged phase velocity tensor in the random medium 
\cite{Tig92,Kro93}, and the energy transport velocity tensor 
${\b v}_{\mbox{\tiny E}} (\omega)$ is to be identified below.
We stress that in Eqs. (\ref{PE}) and  
(\ref{JE}) the products on the r.h.s. are contractions. For 
instance, in full index notation Eq. (\ref{PE}) reads
\begin{eqnarray}
\left[ P^{\omega}_{\mbox{\tiny E}}(\vec{Q}, \Omega) \right]_{\alpha\beta}
 = 
 \omega^2  \left[ c^{-2}_{\mbox{\tiny P}}\right]_{ij} 
 \left[\int\!\!\! \frac{d^3q}{(2 \pi)^3} \,
       {\b \Phi}^{\omega}_{\vec{q}} 
                 (\vec{Q},\Omega)\right]_{i j \alpha \beta}\!, 
\end{eqnarray}
where summation over repeated indices is implied. A corresponding 
expression can be written down for the current density tensor,
Eq. (\ref{JE}). \par

The expressions Eqs.\ (\ref{PE}), (\ref{JE}) 
are the leading terms of a small $\Omega$, $\vec Q$ expansion 
of $\b \Phi$ \cite{Kopp84,Kro90}, analogous to the 
so-called  P$_1$-approximation to the standard Boltzmann 
equation \cite{Cas67},
\begin{eqnarray}
{\b \Phi}^{\omega}_{\vec{q}}({\vec{Q}},\Omega)
& \approx & 
\Delta
{\b G}_{\vec{q}}^{\omega}(0,0)
\left[
\tilde{P}_{E}^{\omega}({\vec{Q}},\Omega){\b A}_0^{-1}
\right. \nonumber \\
& & \left.+ ({\vec{q}}\cdot {\vec{Q}})
\tilde{J}_{E}^{\omega}({\vec{Q}},\Omega){\b A}_1^{-1}
\right] + O(\Omega, \vec Q^2)\\
{\b A}_0 &=& \int d^3 q  \Delta {\b G}^{\omega}_{\vec{q}} (0,0) \\
{\b A}_1 &=& \int d^3 q  (\hat{q}\cdot \hat{Q}) ^2 \Delta {\b G}^{\omega}_{\vec{q}} (0,0),
\end{eqnarray}
The tensor coefficients $\b A _0$ and $\b A _1$ above can be 
computed by projecting ${\b \Phi}^{\omega}_{\vec{q}}({\vec{Q}},\Omega)$
onto its 0th and 1st moments with respect to the 
longitudinal current vertex $(\vec q \cdot \hat Q)$, 
i.e. onto ${\b P}^{\omega}_{\mbox{\tiny E}}(\vec{Q}, \Omega)$
and ${\b J}^{\omega}_{\mbox{\tiny E}}(\vec{Q}, \Omega)$ respectively.

The continuity equation for ${\b P}_{\tiny E}$ and ${\b J}_{\tiny E}$, can
be derived by integrating the generalized Boltzmann equation,
Eq. (\ref{boltzmann}), over the momentum $\vec{q}$ and subsequent
application of the WI, Eq. (\ref{Ward_2}). Upon 
considering the long-time, long-distance limit, we arrive at
\begin{eqnarray}
\label{continuity}
&\!\!\! \!\!\!\!\!\!\!\!\!& -\frac{1}{2} g^{(1)}_{\omega} 
             \left[ \mathbbm{1}\otimes\mathbbm{1} + 
                    {\b \Delta}(\omega) \right] \, 
                    \Omega {\b P}^{\omega}_{\mbox{\tiny E}} + 
                    \omega \left[{\b c}_{\mbox{\tiny P}} \, 
                           {\b v}_{\mbox{\tiny E}}(\omega)\right]^{-1} \,
                           Q {\b J}^{\omega}_{\mbox{\tiny E}} = 
\nonumber\\
&\!\!\!\!\!\!\!\!\!\!\!\!& 
   \omega^2 {\b c}_p^{-2} \!\!\!\int \!\!\!d^3 q  
            \Delta {\b G}^{\omega}_{\vec{q}} (\vec{Q},\Omega) 
             + 
             \frac{1}{2} i \!\! 
              \left[ g^{(0)}_{\omega} \mathbbm{1}\otimes\mathbbm{1}    
                     + {\b\Lambda}(\omega) \right]
              {\b P}^{\omega}_{\mbox{\tiny E}} . 
\end{eqnarray}
The renormalization terms 
${\b \Delta}(\omega)$ and ${\b \Lambda}(\omega)$
originate from the nonzero l.h.s. of the WI, 
Eq. (\ref{Ward_2}), and are defined as
\begin{eqnarray}
\label{delta}
{\b \Delta}(\omega) & = & {\b \Delta}^{(1)}(\omega) + 
                          {\b \Delta}^{(2)}(\omega) \\
\label{lambda}
{\b \Lambda}(\omega) & = & 
\frac{a^{(0)}_{\omega}g^{(1)}_{\omega}}{a^{(1)}_{\omega}}
  {\b \Delta}^{(1)}(\omega),
\end{eqnarray}
with the explicit forms of 
$\Delta^{(1)}(\omega)$ and $\Delta^{(2)}(\omega)$
given by
\begin{eqnarray}
\label{delta1}
\b \Delta^{(1)} (\omega)
&\!\!\!=\!\!\!&-
\frac{{a}^{(1)}_{\omega}}{g^{(1)}_{\omega}}
\!\!\int\!\!\!\frac{{d}^3 q }{(2\pi)^3}
\Bigg\{   \Pi {\b \Sigma}^{\omega}_{{\vec{q}}}(0,0)
\Delta
{\b G}_{\vec{q}}^{\omega}(0,0) \Bigg.
\\\nonumber 
+ \Bigg.&\!\!\! \!\!\!&\!\!\!
\!\!\!\int \!\!\! \frac{{\rm d}^3 q^{\prime} }{(2\pi)^3}
\Pi {\b G}^{\omega}_{{\vec{q}}^{\prime}}(0,0)
{  \b \gamma}^{\omega}_{{\vec{q}}^{\prime}{\vec{q}}}(0,0)
\Delta
{ \b  G}_{\vec{q}}^{\omega}(0,0)
\Bigg\} \b A_0^{-1}
\\
\label{delta2}
\b \Delta^{(2)} (\omega)
&\!\!\!=\!\!\!& - i
\frac{{\rm a}^{(0)}_{\omega}}{g^{(1)}_{\omega}}
\!\!\int\!\!\!\frac{{\rm d}^3 q }{(2\pi)^3}
\Bigg\{\!\!
\left[  \partial_{\Omega}    {\b \Sigma}^{\omega}_{{\vec{q}}}(0,0)\right]
\Delta {\b G}_{\vec{q}}^{\omega}(0,0) \Bigg.
 \\
+
\Bigg.&\!\!\!\!\!\!&\!\!\!
\!\!\!\int\!\!\!\frac{{\rm d}^3 q^{\prime} }{(2\pi)^3}
\!\!\left[  \partial_{\Omega}     {\b G}^{\omega}_{{\vec{q}}^{\prime}}(0,0)
\right]\!\!
{\b \gamma}^{\omega}_{{\vec{q}}^{\prime}{\vec{q}}}(0,0)
\Delta
{\b G}_{\vec{q}}^{\omega}(0,0) \Bigg.
\nonumber \\\nonumber
 +
\Bigg.
\int&\!\!\!\!\!\!&\!\!\!\!\!\!\frac{{\rm d}^3 q^{\prime} }{(2\pi)^3}
    \Pi {\b G}^{\omega}_{{\vec{q}}^{\prime}}(0,0)
\!\!\left[  \partial_{\Omega} { \b\gamma}^{\omega}_{{\vec{q}}^{\prime}{\vec{q}}}(0,0)\right]\!\!
\Delta
{ \b G}_{\vec{q}}^{\omega}(0,0)
\!\Bigg\}
\b A_0^{-1}.
\end{eqnarray}
In Eqs. (\ref{continuity}-\ref{delta2}) we have expanded the prefactor
\begin{equation}
a_{\omega}(\Omega) \equiv 
\frac{h(\omega_+) - h^*(\omega_-)}{h(\omega_+) + h^*(\omega_-)} =
i a^{(0)}_{\omega} +  a^{(1)}_{\omega} \Omega + \cdots ,
\end{equation}
to leading order in $\Omega$, where we introduced 
\begin{eqnarray}
\!\!\!\!\!\!\!\!\!\!a^{(0)}_{\omega} &\!\!\!=\!\!\!& \frac {\mbox{Im} \left[ h(\omega) \right] }       {\mbox{Re} \left[ h(\omega) \right] }    
\\
\!\!\!\!\!\!\!\!\!\!a^{(1)}_{\omega} &\!\!\!=\!\!\!&  \frac {\mbox{Re}\left[ h(\omega)  \right] \mbox{Re}\left[ \partial_{\omega}h(\omega)  \right] + 
                             \mbox{Im} \left[ h(\omega) \right] \mbox{Im}\left[ \partial_{\omega}h(\omega)  \right]}
                         {2 \mbox{Re}\left[ h(\omega) \right]}  .
\end{eqnarray}
Eq. (\ref{continuity}) takes the form of a continuity equation by 
identifying the  energy transport velocity 
${\b v}_{\mbox{\tiny E}} (\omega)$ such that the 
prefactors of the two terms on the l.h.s. are equal, 
and by multiplying with the inverse of that prefactor.
In fact, this procedure defines the energy transport velocity, 
\begin{equation}
\label{velocity}
{\b v}_{\mbox{\tiny E}} (\omega) = 
\frac{\omega}{g^{(1)}_{\omega}} 
       \left[{\b c}_{\mbox{\tiny P}}\right]^{-1}
       \left[ \mathbbm{1}\otimes\mathbbm{1} + 
              {\b \Delta} (\omega)\right]^{-1},
\end{equation} 
where the r.h.s. implies a contraction 
as explained  in Eq. (\ref{PE}). The continuity equation then reads,
\begin{eqnarray}
\label{continuity1}
&\!\!\!\!\!\!\!\!\!\!\!\!&                    \Omega {\b P}^{\omega}_{\mbox{\tiny E}} \,\, + \,\,
                           Q {\b J}^{\omega}_{\mbox{\tiny E}} = 
\nonumber\\
&\!\!\!\!\!\!\!\!\!\!\!\!& 
    -\frac{2}{g^{(1)}_{\omega} }   \left[ \mathbbm{1}\otimes\mathbbm{1} + 
                    {\b \Delta}(\omega) \right]^{-1}    \omega^2 {\b c}_p^{-2} \!\!\!\int \!\!\!d^3 q  
            \Delta {\b G}^{\omega}_{\vec{q}} (\vec{Q},\Omega) \\
&\!\!\!\!\!\!\!\!\!\!\!\!& 
             + 
             \frac{1}{g^{(1)}_{\omega}} i    \left[ \mathbbm{1}\otimes\mathbbm{1} + 
                    {\b \Delta}(\omega) \right]^{-1}    
 \!\! 
              \left[ g^{(0)}_{\omega} \mathbbm{1}\otimes\mathbbm{1}    
                     + {\b\Lambda}(\omega) \right]
              {\b P}^{\omega}_{\mbox{\tiny E}} . \nonumber
\end{eqnarray}
We want to emphasize that in the long-time, long-distance limit,
the continuity equation, Eq. (\ref{continuity1}) is exact.
Also note that in the disordered medium transport occurs, with
different velocities, both perpendicular
and parallel to the polarization directions of the in- as well as
of the outgoing wave field, since each scattering process alters the 
polarization. Accordingly, the averaged transport velocity 
${\b v}_{\mbox{\tiny E}} (\omega)$
is a 4th rank tensor, as evidenced by Eq.\ (\ref{velocity}). 
Similarly, in a transport theory of a vector field 
the diffusion coefficient ${\b D}(\omega)$, the transport mean free path
${\b \ell}_{\mbox{\tiny T}} (\omega)$, the absorption/gain 
length ${\b x}_a^{}(\omega)$, given below by Eqs.\ (\ref{diffusion}), 
(\ref{lt}), and (\ref{massterm}), respectively, and other transport quantities
are tensors as well.

The microscopic derivation of the energy transport velocity,
Eq.\ (\ref{velocity}), and the continuity equation, 
Eq.\ (\ref{continuity1}),
exhibit a number of important physical aspects: 

(1) The frequency dependent
renormalization ${\b \Delta} (\omega)$ of the energy transport 
velocity tensor consists of the ``dwell time'' renomalization
${\b \Delta}^{(1)} (\omega)$ alluded to in the introduction which in the case
of independent scatterers may be traced to their internal (Mie)
resonances. 

(2) However, in the presence of a mismatch in absorption
or amplification (``impedance mismatch'') between the scatterer 
and the host medium (non-vanishing $a_\omega^{(0)}$), Eq.\ (\ref{velocity}) 
features additional renormalization of the transport velocity. To the best
of our knowledge, this renormalization has not been discussed 
before. Numerical results for the transport velocity in 
absorptive or active media will be presented in Section VI.

(3) The r.h.s. of the continuity equation, Eq. (\ref{continuity1})
features as the first term the source contribution 
expected for correlation functions. We find that 
in the presence of either absorption 
or gain the second term on the r.h.s. does not
vanish. Moreover, a ``dwell time'' renormalization 
$ {\b \Lambda} (\omega) \propto {\b \Delta}^{(1)} (\omega)$
occurs in the case of unequal absorption or gain characteristics
of host medium and scatterer (non-vanishing $a_\omega^{(0)}$).
While such a renormalization of either absorption or gain could
be expected on physical grounds, our calculation constitutes the
first quantitative analysis of this effect.
Both absorption/gain related renormalization terms may have
profound influence on lasing action in random dielectric media
and further explorations are necessary.

\section{Fick's law and diffusion pole}
In order to close the set of equations for 
${\b P}_{\mbox{\tiny E}}$ and ${\b J}_{\mbox{\tiny E}}$,
another relation besides the continuity equation
has to derived from the kinetic equation, Eq.\ (\ref{boltzmann}), 
which relates the 
current correlator ${\b J}_{\mbox{\tiny E}}$
to the gradient of the density correlator ${\b P}_{\mbox{\tiny E}}$, 
i.e. a version of Fick's law for the correlation tensors.
This can be achieved by first
multiplying Eq.\ (\ref{boltzmann}) with $\vec{q}\cdot
\hat{Q}$, subsequent integration over the momenta $\vec{q}$ and 
employing the WI, Eq. (\ref{Ward_2}). The resulting
relation should be evaluated to first order in momentum $Q$ at
zero external frequency, $\Omega = 0$ (see also the book by Case
and Zweifel \cite{Cas67} for a detailed discussion of this so-
called P$_1$-approximation). This procedure leads to Fick's law,
\begin{eqnarray} 
\label{ficklaw}
Q \left[{\b A} + {\b \kappa} (\omega) \right]
       {\b P}^{\omega}_{\mbox{\tiny E}}  
   = g_{\omega}^{(1)}{\b \ell}^{-1} 
     \left[\mathbbm{1}\otimes\mathbbm{1} + 
           {\b \Delta}(\omega) \right] {\b J}^{\omega}_{\mbox{\tiny E}},
\end{eqnarray}
where the renormalization terms {\b A} and ${\b \kappa}$ 
are given by 
\begin{eqnarray} 
{{\b A}} &=& \!\!\!\int\!\!\! \frac{{\rm d}^3 q }{(2\pi)^3} 
             ({\vec{q}}\cdot\hat{Q})^2 
             \Delta {\b G}_{\vec{q}}^{\omega}(0,0){\b A}_0^{-1} \\
{\b \kappa}(\omega) 
&=&
ia_{\omega}^{(0)}\!\!\!
\int\!\!\!\frac{{\rm d}^3 q}{(2\pi)^3}
({\vec{q}}\cdot\hat{\vec{Q}})
\partial_{\vec{Q}}
\Big[
{\b\Pi} {\b \Sigma}_{\vec{q}}^{\omega}(0,0)
\\
&\!\!\!+&\!\!\!\!\!\!
\int\!\!\!\frac{{\rm d}^3 q^{\prime} }{(2\pi)^3}
{\b \Pi} {\b G}_{{\vec{q}}^{\prime}}^{\omega}(0,0)
{\b\gamma}_{{\vec{q}}^{\prime}{\vec{q}}}(0,0)
\Big]\!\cdot\!
{\hat{\vec{Q}}}
\Delta {\b G}_{{\vec{q}}}^{\omega}(0,0){\b a}_0^{-1}.
\nonumber
\end{eqnarray}
The tensor quantity $\b{\ell}^{-1}(\omega)$ (of dimension inverse 
length)
\begin{eqnarray} 
\label{transportmeanfreepath}
{\b \ell}^{-1}(\omega)
&=&
 \,\,\, ig_{\omega}^{(0)}
\mathbbm{1}\otimes\mathbbm{1}
 +\Delta_{\omega}^{(3)} + {\b \ell}_0^{-1} (\omega)
\end{eqnarray}
consists of a ``standard'' term
$\b{\ell}^{-1}_0 (\omega)$ that would also be there in 
electron transport theory, 
\begin{eqnarray} 
\nonumber
{\b \ell}_0^{-1} (\omega) &=&   -
\!\!\int\!\!\!\frac{{\rm d}^3 q}{(2\pi)^3}
\!\!\int\!\!\!\frac{{\rm d}^3 q^{\prime} }{(2\pi)^3}
({\vec{q}}\cdot\hat{\vec{Q}})^2
\Delta {\b G}_{{\vec{q}}^{\prime}}^{\omega}(0,0)\\
&&\quad\times
\b\gamma_{{\vec{q}}^{\prime}{\vec{q}}}(0,0)
\Delta {\b G}_{{\vec{q}}}^{\omega}(0,0){\b a}_0^{-1}
\nonumber\\\nonumber
&&+
\!\!\int\!\!\!\frac{{\rm d}^3 q}{(2\pi)^3}
\!\!\!\int\!\!\!\frac{{\rm d}^3 q^{\prime} }{(2\pi)^3}
({\vec{q}}\cdot\hat{\vec{Q}})({\vec{q}}^{\prime}\cdot\hat{\vec{Q}})
\Delta {\b G}_{{\vec{q}}}^{\omega}(0,0)\\
&&\quad\times
\b\gamma_{{\vec{q}}{\vec{q}}^{\prime}}(0,0)
\Delta {\b G}_{{\vec{q}}^{\prime}}^{\omega}(0,0){\b a}_1^{-1}, 
\end{eqnarray}
and two additional contributions 
$g^{(0)}_\omega \mathbbm{1}\otimes\mathbbm{1}$ and
${\b \Delta}_{\omega}^{(3)}$ that are non-vanishing only in the presence 
of either absorption or gain,
\begin{eqnarray} 
{\b \Delta}_{\omega}^{(3)}
&=&
-i A_{\omega}^{(0)}
\!\!\!\int\!\!\!\frac{{\rm d}^3 q}{(2\pi)^3}
({\vec{q}}\cdot\hat{\vec{Q}})^2
\Big[
\b\Pi {\b \Sigma}_{\vec{q}}^{\omega}(0,0)
\\
&&\!\!\!\!\!\!
+
\!\!\int\!\!\!\frac{{\rm d}^3 q^{\prime} }{(2\pi)^3}
\b\Pi {\b G}_{{\vec{q}}^{\prime}}^{\omega}(0,0)
\b\gamma_{{\vec{q}}^{\prime}{\vec{q}}}(0,0)
\Big]\Delta {\b G}_{{\vec{q}}}^{\omega}(0,0){\b A}_1^{-1}.
\nonumber
\end{eqnarray}
Similar to the case of the energy transport velocity,
any discrepancy between the absorption/gain characteristics 
of scatterers and host medium (``impedance mismatch'' given
by a non-vanishing $a^{(0)}_{\omega}$) leads to a renormalization 
of ${\b \ell}^{-1}$. Although the renormalizations 
${\b \Delta}^{(3)} (\omega)$ and $\b\kappa (\omega)$ originate
from isotropic and anisotropic scattering effects, respectively, 
they contribute only in the presence of absorption/gain mismatch 
between scatterer and host medium. This is in contrast to earlier 
results by Livdan and Lisyansky \cite{Liv96} for media with 
frequency independent absorbing spheres, due to the fact that these 
authors inconsistently employed the WI without 
absorption/gain when deriving the current-density relation 
Eq. (\ref{ficklaw}).\par
Combining the continuity equation, Eq. (\ref{continuity}), with
the microscopic version of Fick's law, Eq. (\ref{ficklaw}), we are
finally in a position to solve for the energy density tensor in the
long-time, long-distance limit
\begin{eqnarray}
\label{diffusionpole}
{P}_{\mbox{\tiny E}}^{\omega}
&\!\!\!=\!\!\!&
\left(\!\!
-\frac i2 g_{\omega}^{(1)} \left( 1+\Delta (\omega) \right)
\left( -i\Omega + {\b D} \left( Q^2 + {\b x}_a^{-2} \right) \right) \!
\right)^{-1} \nonumber
\\
&&\qquad\times
\omega^2\b c_p^{-2}\int\!\! d^3 q  
\Delta {\b G}^{\omega}_{\vec{q}} (0,0) 
\end{eqnarray}
which exhibits a familiar diffusion pole structure. In addition, 
the presence of absorption or gain leads to the appearance of a 
mass term ${\b D}(\omega)/{\b x}^{-2}_a(\omega)$.
This term is absent in the work of Livdan and Lisyansky 
\cite{Liv96}, although they considered absorbing spheres, 
albeit within a scalar model. 
From Eq.\ (\ref{diffusionpole}) 
the diffusion tensor ${\b D}(\omega)$ is explicitly given by
\begin{eqnarray}
\label{diffusion} 
 {\b D}(\omega) & \!\!=\!\! & 
\frac{2i}{g^{(1)}_{\omega}} 
\left( \mathbbm{1}\otimes\mathbbm{1} + {{\b \Delta} (\omega)} \right)^{-1} 
\b\ell \left( {{\b A}} + {{\b\kappa} (\omega)} \right) \\
&\!\! \equiv\!\! & \frac{1}{3} \, {\b v}_{\mbox{\tiny E}} (\omega) \, 
                          {\b \ell}_{\mbox{\tiny T}} (\omega) \nonumber ,
\end{eqnarray}
which establishes the transport mean free path tensor 
${\b \ell}_{\mbox{\tiny T}} (\omega)$ as
\begin{equation}
\label{lt}
{\b \ell}_{\mbox{\tiny T}} (\omega) = 
3 \frac{2i} {\omega} {\b c}_{\mbox{\tiny P}} {\b \ell}\left( {{\b A}} + {{\b\kappa} (\omega)} \right)
\end{equation}
Finally, the absorption/gain length tensor ${\b x}_a^{}(\omega)$
explicitly reads 
\begin{equation} 
\label{massterm}
{\b x}_a^{2}
= \frac 13 {\b\ell}_A{\b\ell}_T, 
\end{equation}
where the absorption mean free path tensor $\b\ell_A$ is given by
\begin{equation} 
{\b\ell}_A = \frac {\omega}{g^{(0)}_{\omega}}
\left( \mathbbm{1}\otimes\mathbbm{1} + {{\b \Lambda} (\omega)} \right)^{-1}.
\end{equation}
Eqs. (\ref{velocity}), (\ref{diffusion}), (\ref{lt}), and (\ref{massterm})
represent the central results of our investigation and make explicit how the
various renormalizations discussed above affect the transport quantities
of a disordered dielectric medium in the presence of linear absorption
or gain. In the following section, we provide illustrations of these
results for a system of dilute scatterers.

\begin{figure}[t]
\begin{center}
\scalebox{0.33}{\includegraphics{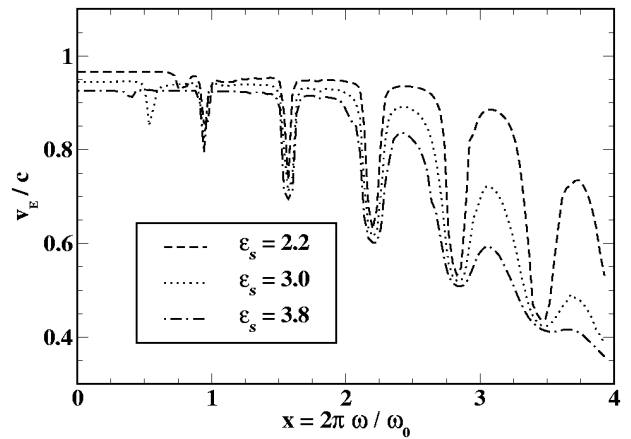}}
\end{center}
\caption{Frequency dependence of the energy transport velocity 
$v_{\mbox{\tiny E}}$ units of the vacuum speed of light for a dilute
($n = 6 \% $ by volume) collection of spherical scatterers in air 
($ \epsilon_h = 1$) with different dielectric constants $\epsilon_s$ 
(as indicated). 
The frequency unit is $\omega_{\mbox{\tiny 0}} = 2\pi c/r_{\mbox{\tiny 0}}$ 
where $c$ is the vacuum speed of light
and $r_{\mbox{\tiny 0}}$ the radius of the scatterers. 
The pronounced dips correspond 
to the Mie resonances of the individual sphere and represent the ``dwell
time'' effect discussed in the text.}
\label{plot1}
\end{figure}
\section{Numerical results}
As alluded to above, any numerical evaluation of 
Eqs. (\ref{velocity}), (\ref{diffusion}), (\ref{lt}), and (\ref{massterm})
requires the computation of consistent values for the self-energy tensor
${\b \Sigma}^{\omega_{\pm}}_{\vec{q}_{\pm}}$ 
and the irreducible vertex tensor
${\b \gamma}^{\omega}_{\vec{q} \vec{q}^\prime}(\vec{Q},\Omega)$.  
For dense systems and in the diffusive regime, a Coherent Potential 
Approximation has to be employed for both, the self-energy tensor and 
the irreducible vertex tensor. To date, such an approach has been
carried out for scalar waves and in the presence of point scatterers 
only \cite{Gon92,Kro93}. Moreover, near the Anderson localization 
transition, the irreducible vertex tensor is expected to exhibit  
critical behavior and more sophisticated schemes such as a 
self-consistent theory of localization  
have to be employed. Again, to date such a program 
has been realized for scaler waves and point scatterers only 
\cite{Vol80a,Vol80b,Kro93}.\par
However, in realistic strongly scattering electromagnetic systems, the
scatterers cannot be approximated by point scatterers. Instead, they
exhibit internal (Mie) resonances which - besides the vectorial nature
of the electromagnetic radiation itself - greatly complicate
the approach discussed above. Therefore, we have chosen to illustrate
our results in a simpler system consisting of a dilute collection of
identical absorbing or amplifying spherical scatterers in a host medium 
without absorption or gain. In this case, the independent scatterer
approximation applies and the self-energy as well as the irreducible
vertex tensor may be expressed through the density $n$ of scatterers
and the full t-matrix of such a scatterer according to \cite{Tig93,Gon92}
\begin{eqnarray}
{\b \Sigma}^{\omega_{\pm}}_{\vec{q}_{\pm}}
&=&
n \;{\b t}_{ {\vec{q}}_{\pm},{\vec{q}}_{\pm} } (\omega_{\pm}) \\
{\b \gamma}_{ {\vec{q}}, {\vec{q}}^\prime } (\vec{Q};\Omega)
&=&
n \;
{\b t}_{ \vec{q}_{+},\vec{q}^\prime_{+} } (\omega_{+})
\otimes
{\b t}_{ \vec{q}^\prime_{-},\vec{q}_{-} } (\omega_{-}).
\end{eqnarray}
Explicit expressions for the t-matrix of a spherical scatterers are
fairly involved and can be found in a recent work by K. Arya {\em et al.} 
\cite{t-matrix-ref}. Consequently, even within this approximation,
all transport quantities have to be evaluated numerically.
 
\begin{figure}[b] 
\begin{center}
\rotatebox{0}{\scalebox{0.33}{\includegraphics{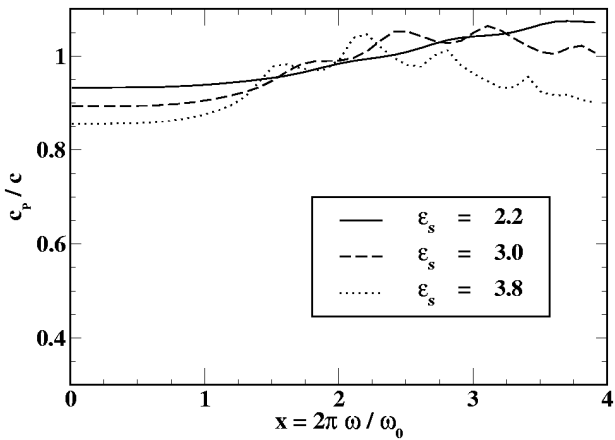}}}
\end{center}
\caption{Frequency dependence of the phase velocity 
$c_{\mbox{\tiny P}}$ units of the vacuum speed of light for a dilute
($n = 6 \% $ by volume) collection of spherical scatterers in air 
($ \epsilon_h = 1$) with different dielectric constants $\epsilon_s$ 
(as indicated). 
The frequency unit is $\omega_{\mbox{\tiny 0}} = 2\pi c/r_{\mbox{\tiny 0}}$ 
where $c$ is the vacuum speed of light
and $r_{\mbox{\tiny 0}}$ the radius of the scatterers.
Near Mie resonances, this
velocity significantly exceeds the vacuum speed of light. The corresponding
physical energy transport velocity is displayed in Fig. \ref{plot1}.}
\label{plot1a}
\end{figure}
Note that the integrations with respect to momentum of, e.g.,  
certain convolutions of the irreducible Vertex $\b \gamma$, which 
occur in the evaluation of the transport quantities, are in general
ultraviolet divergent. These divergencies
are remedied by applying the regularization scheme familiar 
from quantum electrodynamics \cite{Itz78}, i.e. by subtracting 
appropriate mass terms from the integrands such that 
after subtraction, these terms,
and sufficiently many of their derivatives, do not diverge, 
thus making the integrals convergent. Note that the 
corresponding regularization masses in general 
have to be determined numerically, since the prefactors of the
ultraviolet divergencies are known numerically only.

\begin{figure}[t]
\begin{center}
\scalebox{0.33}{\includegraphics{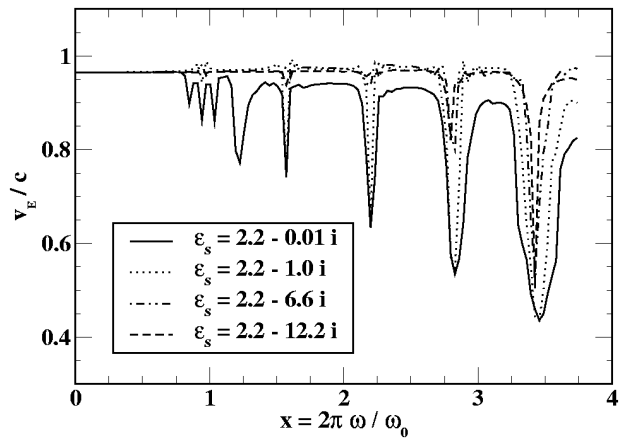}}
\end{center}
\caption{Frequency dependence of the energy transport velocity 
$v_{\mbox{\tiny E}}$ units of the vacuum speed of light for a dilute
($n = 6 \% $ by volume) collection of spherical scatterers in air 
($ \epsilon_h = 1$) where the scatterer dielectric constant $\epsilon_s$
exhibits different amounts of linear  gain  (as indicated). 
The frequency unit is $\omega_{\mbox{\tiny 0}} = 2\pi c/r_{\mbox{\tiny 0}}$ 
where $c$ is the vacuum speed of light
and $r_{\mbox{\tiny 0}}$ the radius of the scatterers.
 For increasing gain, the dips
in the energy transport velocity that are associated with the individual 
scatterers' Mie resonances exhibit considerable narrowing. This may be
interpreted as a sign for the incipient onset of random lasing action
in this system.}
\label{plot2}
\end{figure}
In the following we restrict ourselves  
to a discussion of the numerical results for the 
energy transport velocity tensor. 
We would like to point out that for isotropic dielectric 
materials, the energy transport velocity tensor becomes an isotropic
tensor, so that we can further restrict our discussion to a single
valued, frequency dependent energy velocity.\par
In Fig. \ref{plot1}, we display the frequency dependence of the 
energy transport velocity for a typical concentration of 6 \% by 
volume of absorption-free scatterers with different (real) dielectric 
constants in an air background. Clearly visible are the pronounced
dips in the energy transport velocity near frequencies that correspond
to the single scatterer Mie resonances. As the dielectric constant 
increases, the Mie scattering becomes stronger and leads to 
correspondingly
stronger renormalization of the energy transport velocity through the
``dwell time'' effect mentioned above. These data are, therefore, 
consistent
with earlier results for scalar waves \cite{Alb91,Tig93} and vector waves
\cite{Bus95,Bus96}.
This behavior of the energy transport velocity has to be contrasted 
with the corresponding frequency dependence of the phase velocity, 
Eq. (\ref{phasevelocity}) in Fig. \ref{plot1a}. Near the Mie resonances,
this phase velocity may exhibit values that significantly exceed the 
vacuum speed of light. Since the phase velocity in a random medium 
does not correspond to a physical quantity, this behavior is not
in conflict with the laws of relativity. \par
Adding linear gain to the system, dramatically modifies the situation.
In Fig. \ref{plot2}, we show the frequency dependence of the energy 
transport velocity for the weakest scattering system of Fig. \ref{plot1}
linear gain is added to the scatterer. Although the concentration is
only $n = 6 \%$ by volume, adding a relatively small negative imaginary 
part to the dielectric constant results in a significant narrowing of 
the energy transport velocity resonances. This gain narrowing 
may be understood by noting that the Mie resonance dips in the 
transport velocity arise because of multiple reflection and 
interference of light 
within a single scatterer (``dwell time effect''). In the presence of
gain, the relative importance of long wave paths for the 
interference processes is enhanced, leading to a narrowing of the
resonance lines, analogous to the narrowing of the transmission lines
in e Fabry-Perrot filter with increasing number of reflections. 
Increasing the gain to rather unrealistic values shows further narrowing
of the resonances together with a slight increase in the energy transport 
velocity. It should be noted that the energy transport velocity retains
physical values for all values of the gain that we have considered. \par
\begin{figure}
\begin{center}
\scalebox{0.33}{ \includegraphics{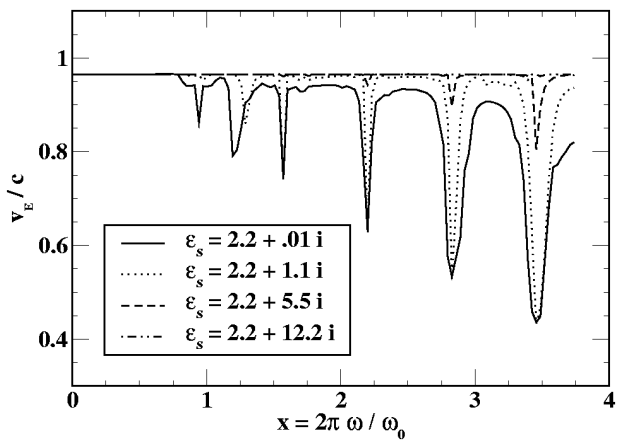}}
\end{center}
\caption{Frequency dependence of the energy transport velocity 
$v_{\mbox{\tiny E}}$ units of the vacuum speed of light for a dilute
($n = 6 \% $ by volume) collection of spherical scatterers in air 
($ \epsilon_h = 1$) where the scatterer dielectric constant $\epsilon_s$
exhibits different amounts of linear absorption (as indicated). 
The frequency unit is $\omega_{\mbox{\tiny 0}} = 2\pi c/r_{\mbox{\tiny 0}}$ 
where $c$ is the vacuum speed of light
and $r_{\mbox{\tiny 0}}$ the radius of the scatterers.
For increasing absorption, the 
dips in the energy transport velocity that are associated with the individual 
scatterers' Mie resonances become washed out and disappear altogether
for sufficiently large absorption, as the wave propagation becomes
overdamped and interference effects become impossible.}
\label{plot3}
\end{figure}
The opposite behavior occurs when we add linear absorption to the weakest
scattering spheres of Fig. \ref{plot1}. This is illustrated in Fig. 
\ref{plot3}, where we display the frequency dependence of the energy 
transport for increasing values of linear absorption in this system. 
In this case, the dips in the energy transport velocity that are 
associated with the Mie resonances of the individual scatterers become 
washed out and disappear altogether for sufficiently large values of 
the linear absorption.

\section{Conclusion}
In conclusion, we have presented a microscopic approach to the calculation
of transport quantities for electromagnetic waves propagating in disordered
dielectric media which exhibit linear absorption or gain. The effects 
of energy conservation and of the violation of energy conservation
in the presence of absorption or gain have been incorporated in the theory by 
means of an exact Ward identity. The kinetic equation for light in 
random media with absorption or gain has been derived and solved
for the energy density correlation tensor,
taking fully into account the vectorial nature of electromagnetic 
radiation.  In this way we have, 
to the best of our knowledge, for the first time derived explicit 
expressions for the energy transport velocity tensor,
the diffusion tensor, the mean free path tensor, and the absorption length
tensor in absorbing/emitting media.
All these quantities experience renormalization due to scattering or
impedence mismatch between host medium and scatterer material.\par
Specifically, we have discussed the case of a dilute system of identical
spherical scatterers within the framework of the independent scatterer
approximation. These systems exhibit considerable  dips in the energy 
transport velocity that can directly be traced to a well-known ``dwell 
time'' effect.
However, adding linear gain to the scatterers dramatically modifies this
situation and significant gain narrowing occurs already for relatively
modest values of the scatterers concentration and the imaginary part of
their dielectric constant. We interpret this behavior as due to an 
increase of the relative importance of wave paths with long path
lengths in the medium, and a subsequent, gain-induced enhancement of 
interference effects, analogous to the Fabry-Perrot effect.
The opposite effect occurs when
absorption is added to scatterers and the resonances in the energy transport
velocity are washed out and ultimately disappear altogether. In all cases,
the energy transport velocity retains physical values, in contrast to the
phase velocity.\par
As a systemiatic transport theory, expressing all physical quantities
ultimately in terms of the single-photon selfenergy and the 
two-photon irreducible scattering vertex,
our approach may be generalized to include interference effects
like weak or strong localization, e.g.  
in the spirit of the self-consistent theory
of Anderson localization that is well-known for electronic systems.
Together with a replacement of the linear gain by a direct coupling of 
transport equations to the semiclassical laser rate equations this may provide
a microscopic theory for the interplay of optical gain and Anderson
localization of light. Similarly, the theory may be extended to
include optically anisotropic materials such as scatterers immersed
in a liquid crystal or Faraday-active scatterers. 
\section{Acknowledgments}
We thank C.M. Soukoulis for stimulating discussions.
This project was supported in part by Deutsche Forschungsgemeinschaft 
(DFG) through Research Unit 557 grant KR 1726/2 (A.L., J.K.) and by 
grant KR 1726/3 (J.K.) as well as by the Emmy-Noether program of 
the DFG through grant Bu 1107/2-3 (K.B.).

\end{document}